# Creating and probing wide-bandgap nanoribbon-like structures in a continuous metallic graphene sheet

Si-Yu Li[1,§], Mei Zhou[2,§], Jia-Bin Qiao[1,§], Wenhui Duan[2,*], and Lin He[1,*]

[1]Center for Advanced Quantum Studies, Department of Physics, Beijing Normal University, Beijing, 100875, People's Republic of China

[2]State Key Laboratory of Low-Dimensional Quantum Physics and Collaborative Innovation Center of Quantum Matter, Department of Physics, Tsinghua University, Beijing, 100084, People's Republic of China

[§]These authors contributed equally to this work.
* Email: helin@bnu.edu.cn and dwh@phys.tsinghua.edu.cn

**The light-like dispersion of graphene monolayer results in many novel electronic properties in it[1], however, this gapless feature also limits the applications of graphene monolayer in digital electronics[2]. A rare working solution to generate a moderate bandgap in graphene monolayer is to cut it into one-dimensional (1D) nanometre-wide ribbons[3-13]. Here we show that a wide bandgap can be created in a unique 1D strained structure, i.e., graphene-nanoribbon-like (GNR-like) structure, of a continuous graphene sheet via strong interaction between graphene and the metal substrate, instead of cutting graphene monolayer. The GNR-like structures with width of only a few nanometers are observed in a continuous graphene sheet grown on Rh foil by using thermal strain engineering. Spatially-resolved scanning tunnelling spectroscopy revealed bandgap opening of a few hundreds meV in the GNR-like structure in an otherwise continuous metallic graphene sheet, directly demonstrating the realization of a metallic-semiconducting-metallic junction entirely in a graphene monolayer. We also show that it is possible to tailor the structure and electronic properties of the GNR-like structure by using scanning tunnelling microscope.**

The atomically thin structure of graphene monolayer allows the existence of a variety of mechanisms that can tune its electronic properties efficiently[14,15]. Generally, the supporting substrates could affect the electronic properties of graphene in two different aspects. In one aspect, interactions between graphene and the supporting substrates could change the electronic structures of this unique one-atom-thick membrane dramatically and result in novel electronic properties not found in graphene monolayer[15-20]. In the other aspect, the substrates could introduce strain in graphene and lead to spatially varying lattice distortion in it[21-28], which also affects the behaviors of Dirac fermions in graphene. Here, we study electronic properties of a unique 1D strained graphene structure, i.e., the zigzag-GNR-like or chiral zigzag-GNR-like structure (Figure 1), in a continuous graphene sheet grown on Rh foil (see Methods for more details)[17,23]. The GNR-like structures show a wide bandgap of a few hundreds meV and our result demonstrates that the strong interaction between the strained graphene structure and the substrate plays a vital role in determining the electronic properties of the GNR-like structures.

Because of mismatch of thermal expansion coefficients between graphene and the metallic substrates, 1D quasi-period rippled or wrinkled regions are easy to be observed in graphene grown on Rh foils for strain relaxation (see supplemental data Fig. S1)[23,27,28]. In Fig. 1, we show scanning tunnelling microscopy (STM) images of a new and unique 1D graphene structure, which has never been explored before, to relax the in-plane compressive strains induced by the substrate. There is a quite large structural bending occurring only in small regions of the graphene sheet, keeping the lattice of the remaining graphene sheet almost unaffected (Fig. 1b and 1c). Such a unique structure, as shown in Fig. 1, should be energetically unfavorable with comparison to the rippled or wrinkled graphene structures, however, the enhanced interaction between the bending graphene region and the substrate may stabilize it and enable us to investigate its electronic structure. Similar 1D strained structures with different widths $W$ of the flat regions are observed in our STM studies (see supplemental data Fig. S2). In our experiment, it is interesting to note that the bending almost occurs along the zigzag direction of graphene (Fig. 1d, Fig. 1e and Fig. S2) and the angles between the bending direction and the zigzag edge of graphene are measured to be less than 5.5° for all the studied samples. Therefore, the coexistence of the in-plane compressive strains and the interaction between graphene and the Rh foil generates the zigzag-GNR-like (or chiral zigzag-GNR-like) structures in a continuous graphene sheet.

For flat graphene monolayer on a (111) surface of Rh with the equilibrium separation ~ 0.22 nm, the lattice mismatch between graphene and Rh(111) will result in a hexagonal moiré pattern with the expected periodicity of about 2.9 nm[17] (see Fig. S3). In our experiment, we cannot detect any moiré superstructure in the flat regions around the 1D strained structure (Fig. 1 and Fig. S2), indicating that the distance between the flat regions and the substrate is much larger than the equilibrium separation. The distance between the bending graphene region of the 1D strained structure and the substrate is much smaller than that between the flat regions and the Rh surface. Consequently, the π electrons of graphene in the bending region are strongly hybridized with the *d* electrons of the Rh foil, while the electrons in the flat graphene region are almost not affected by the Rh foil due to the relatively large distance between them (in our experiment, the distance between the flat regions and the substrate is estimated to be larger than 0.30 nm according to the profiles of the GNR-like structures). The spatial variation of the interaction with the substrate is expected to change the electronic properties of graphene dramatically[13,29], implying that the (chiral) zigzag-GNR-like structure may exhibit novel electronic properties not found in graphene monolayer.

It has been well established that quantum confinement of charge carriers could result in gap opening in armchair GNRs but not in zigzag (and chiral zigzag) GNRs[1,3-11,30]. To open a gap in zigzag (or chiral zigzag) GNRs, electron-electron interactions are necessary to be taken into account[1,6,8-10,30]. In such a case, the zigzag (or chiral zigzag) GNR features a gap with spin-polarized edge states at each edge and the spin polarizations are in opposite directions in the two edges of a GNR. Consequently, the gap opening of a zigzag (or chiral zigzag) GNR is clearly a signature of the emergence of magnetic order in it, which ensures the successful detecting of edge magnetism on individual GNR by investigating its electronic structure via STM measurements[8-10]. Here we use spatially resolved scanning tunnelling spectroscopy (STS), which reflects the energy-resolved local density of states (LDOS), to study the electronic properties of the (chiral) zigzag-GNR-like structures on Rh foils. Figure 2a shows a representative STM image of a (chiral) zigzag-GNR-like structure with $W \sim 6.2$ nm. There is no similar strained structure flanking the GNR-like structure within about 50 nm. The spectra recorded in the (chiral) zigzag-GNR-like structure reveal a clear bandgap ranging from about 400 meV to 800 meV, whereas, the tunnelling spectra acquired outside of the (chiral) zigzag-GNR-like structure indicate a metallic behavior (Fig. 2b). This result demonstrates explicitly the realization of a metallic-semiconducting-metallic junction[11,13] entirely in a continuous graphene sheet. Similar results have been observed in other different GNR-like structures in our experiment with different STM tips and the spatial variation of the spectra around the GNR-like structures, as shown in Fig. 2b as an example, further removes any possible artificial effects of the STM tips. In the experiment, we find that the energy gap varies a lot even around two opposite edges of the single zigzag-GNR-like structure (Fig. 2b). The edge with a deeper bending region exhibits a much larger gap, which indicates that the interaction strength between the bending region of graphene and the Rh foil is critical in determining the value of the energy gap. The gap opening has also been observed in other (chiral) zigzag-GNR-like structures with different widths, as shown in Fig. 2c, indicating that the existence of energy gap is a robust property of the (chiral) zigzag-GNR-like structure on Rh foils. Moreover, the observed energy gaps in the zigzag-GNR-like structures are on the order of magnitude of the expected gaps for free-standing zigzag GNRs with similar widths[9,10], as shown in Fig. 2d.

To further explore the origin of the energy gap observed in our experiment, we carried out first principles calculations on the electronic structure of the zigzag-GNR-like structure on the Rh(111) surface (see Methods for details). Fig. 3a and Fig. 3b show the top and side view of a zigzag-GNR-like structure with width $W$ on the Rh(111) surface, respectively. The distance between the bending graphene region and the Rh(111) surface is defined as $d$, which is used to describe the interaction strength between graphene and the Rh(111) surface (the interaction strength is expected to increase with decreasing $d$). The zigzag-GNR-like structure shows similar electronic structure to that of pristine graphene if we neglect the interaction between the bending graphene and the substrate, i.e., for sufficiently large $d$ (see supplemental data Fig. S4). The strong interaction between the bending graphene region and the Rh(111) surface affects the electronic structure of the zigzag-GNR-like structure and generates a wide energy gap in it. Fig. 3c shows the electronic structure of the zigzag-GNR-like structure with $W = 1.822$ nm and $d = 0.20$ nm. The obtained energy gap of the zigzag-GNR-like structure is as large as 707 meV, which agrees with our experimental observation reasonably. Previously, it was believed that a moderate energy gap can only be implemented in nanometer-sized GNRs, and no one actually expected it to exist in partial regions of a continuous graphene monolayer. Additionally, for the nanometer-sized GNRs with zigzag edges, it was also believed that electron-electron interactions are vital for the emergence of energy gap and the accompanying magnetic order. Our calculation indicates that the strong interaction between the bending graphene region and the Rh(111) surface is sufficient to result in the gap opening in the GNR-like structures, as observed in our experiment (Fig. 2). Moreover, zigzag-GNR-like structures with antiferromagnetic and ferromagnetic inter-edge couplings are almost degenerate in energy and have very similar electronic structures, indicating that the studied zigzag-GNR-like structure should not exhibit any magnetic order. Therefore, our result reveals a new method to generate a wide bandgap in graphene sheet. The observation of wide energy gaps in the GNR-like structures opens the way towards the tailoring of electronic structures of a continuous graphene sheet by using both the "strain engineering" and the interaction between graphene and the substrates.

Both our experimental and theoretical results indicate that the value of the bandgap in the zigzag-GNR-like structure is mainly affected by two factors, i.e., the width of the zigzag-GNR-like structure $W$ and the distance between the bending structure and the Rh(111) surface $d$. Our calculations point out that the bandgap decreases with increasing width $W$ (for example, it decreases from $E_g$ = 855 meV for a zigzag-GNR-like structure with $W$ = 1.414 nm to $E_g$ = 626 meV for the structure with $W$ = 2.263 nm), whereas, it increases with decreasing distance $d$ [for example, it increases from $E_g$ = 609 meV for the (1.822 nm)-wide zigzag-GNR-like structure with $d$ = 0.220 nm to $E_g$ = 751 meV for the same structure with $d$ = 0.180 nm] (see Fig. S5). In our experiment, the distance $d$ varies dramatically in different zigzag-GNR-like structures. Therefore, there is a wide distribution of the energy gaps obtained in different (chiral) zigzag-GNR-like structures (Fig. 2d). The observed energy gaps of zigzag GNRs on a Au (111) substrate[10] and on a SiC (0001) substrate[9] are also plotted as a function of the width in Fig. 2d, which are overall consistent with our experimental result, especially for the GNR-like structures with $W$ < 4.5 nm. The difference between the energy gaps observed in our experiment and that reported in literature is mainly attributed to the different mechanisms for the gap opening in the 1D structures (the effects of the supporting substrates due to different edge-substrate hybridization may account for the different results reported in refs. 9 and 10). In our experiment, the depth of the bending region, which determines the interaction between graphene and Rh foil, plays a critical role in determining the value of gaps. For some of the (chiral) zigzag-GNR-like structures with $W > 4.5$ nm, the depths of the bending regions are quite large (see supplemental data Fig. S6). Consequently, the observed energy gaps differ much for the (chiral) zigzag-GNR-like structures with almost the same width (Fig. 2d).

Finally, we show the possible ability in tailoring the structure and electronic properties of the GNR-like structures by using STM. During the STM measurements, some of the GNR-like structures become unstable. Fig. 4 shows a representative experimental result: the width of the flat region and the depth of the bending region of a GNR-like structure change dramatically when we measured its STM images. Although this behavior remains to be understood, it is likely that the mechanical force induced by the sub-nanometre distance between the tip and the GNR-like structure plays an important role. The electronic properties of the GNR-like structure also change (Fig. 4b and 4d), accompanied by the variation of its structure. For the structure shown in Fig. 4c, the van Hove singularities induced by 1D electron confinement still persist, whereas, the bandgap-like feature disappears (Fig. 4d), which is attributed to the reduced coupling between the bending graphene and the Rh(111) surface.

chiral graphene nanoribbons. *Phys. Rev. B* **84**, 115406 (2011).

**Acknowledgements**

This work was supported by the National Basic Research Program of China (Grants Nos. 2014CB920903, 2013CBA01603), the National Natural Science Foundation of China (Grant Nos. 11422430, 11374035, 11334006), the program for New Century Excellent Talents in University of the Ministry of Education of China (Grant No. NCET-13-0054), Beijing Higher Education Young Elite Teacher Project (Grant No. YETP0238). L.H. also acknowledges support from the National Program for Support of Top-notch Young Professionals. The first principles calculations were performed on the "Explorer 100" cluster system at Tsinghua University.

**Author contributions**

L.H. conceived and provided advice on the experiment and analysis. W.D. conceived and provided advice on DFT calculation. S.Y.L. and J.B.Q. performed the experiments and analyzed the data. M.Z. performed the theoretical calculations. L.H. wrote the paper. All authors participated in the data discussion.

**Competing financial interests:** The authors declare no competing financial interests.

**Figure Legends**

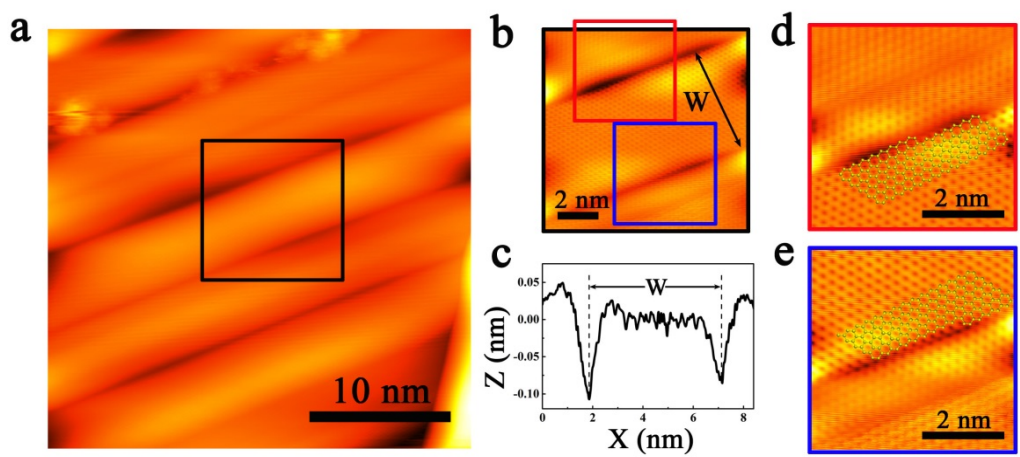

**Figure 1. A (chiral) zigzag-GNR-like structure in a continuous graphene sheet on a Rh foil. a.** A 30 nm × 30 nm STM image of a graphene monolayer ($V_{sample}$ = 800 mV and $I$ = 0.3 nA) on a Rh foil. **b.** A 10 nm × 10 nm STM image of the graphene area within the black frame in **a**. To relax the in-plane compressive strains induced by the substrate, a (chiral) zigzag-GNR-like structure with width of about 5.2 nm is formed. **c.** The profile line of the (chiral) zigzag-GNR-like structure showing a flat region between the two bending boundaries, which are almost parallel. **d** and **e** show atomic resolution STM images of the two bending boundaries of the (chiral) zigzag-GNR-like structure taken from the area marked with the red and blue frames in **b**, respectively. The atomic structures of graphene are overlaid onto the STM images. Obviously, the bending boundaries are almost along the zigzag direction of graphene.

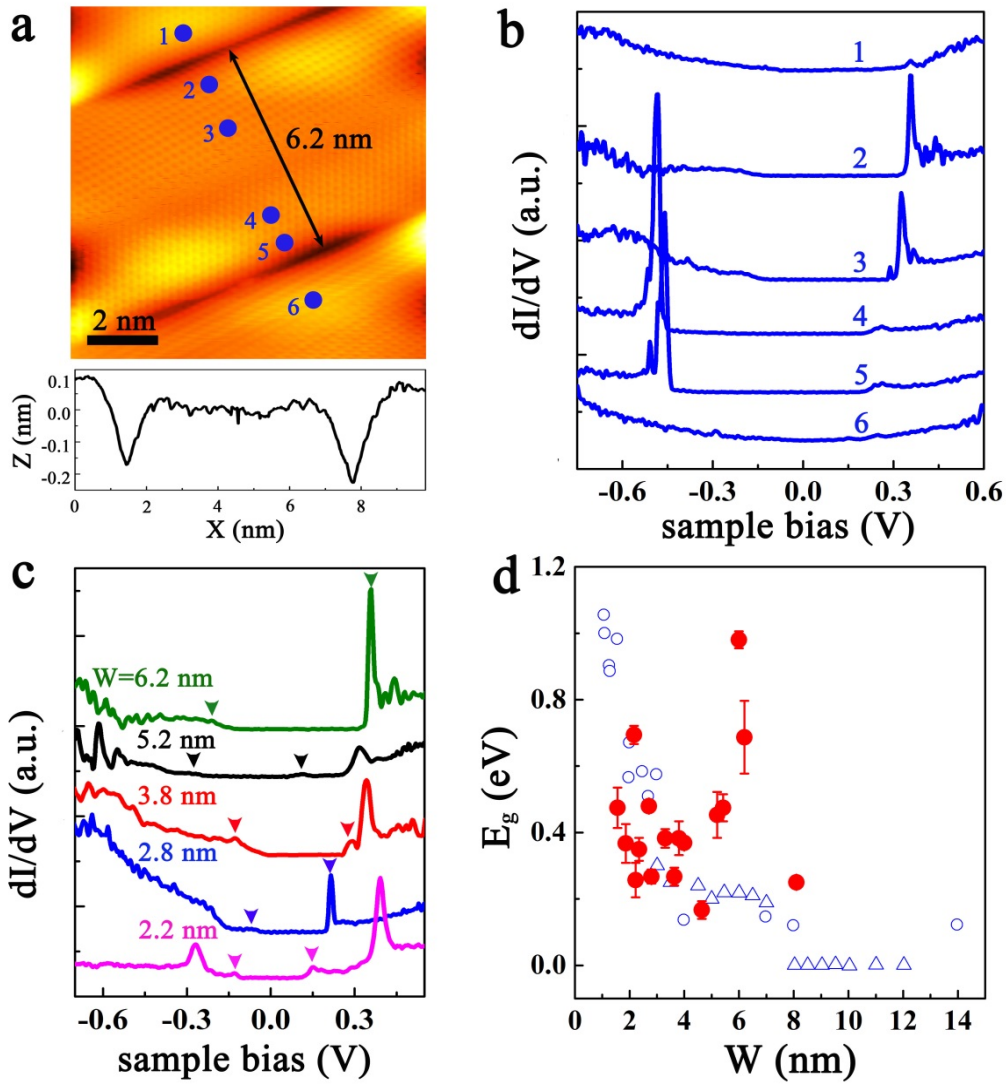

**Figure 2. Energy gaps in the (chiral) zigzag-GNR-like structures. a.** A typical STM image ($V_{sample}$ = 800 mV and $I$ = 0.3 nA) and profile line of a (chiral) zigzag-GNR-like structure with width $W$ ~ 6.2 nm in a continuous graphene layer on the Rh foil. **b.** Spatially resolved STS spectra recorded at different positions from 1 to 6 in **a**. **c.** Typical STS spectra taken on (chiral) zigzag-GNR-like structures with different widths and these samples exhibit different energy gaps, as marked by two arrows. **d.** The red solid symbols show the observed energy gaps as a function of the width in the (chiral) zigzag-GNR-like structures. The open symbols are the experimental data reported in literature very recently: blue open triangles are data

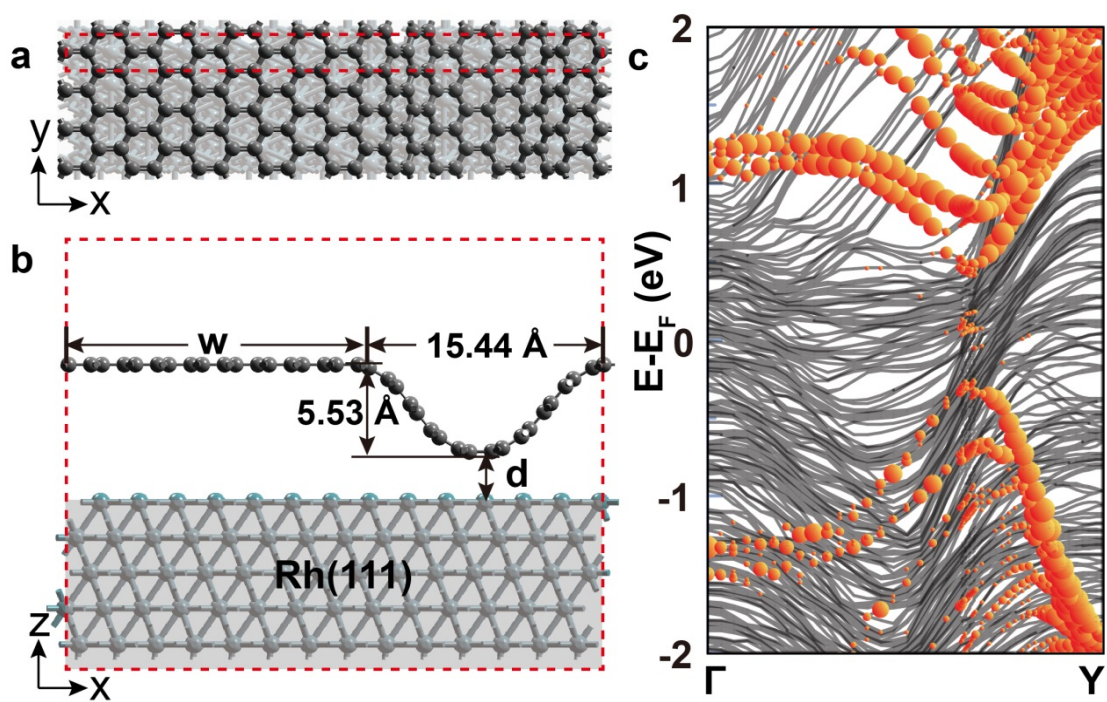

**Figure 3. Geometry and electronic structure of a zigzag-GNR-like structure on Rh(111) surface. a** and **b**: The top and side view of the atomic structure of a zigzag-GNR-like structure on Rh (111) surface (Grey balls: Carbon atoms, blue balls: Rhodium atoms). *W* denotes the width of the flat graphene region and *d* denotes the distance between bending graphene region and the Rh(111) surface. Red dashed frame indicates the supercell in the calculation. **c**. Band structure of the zigzag-GNR-like structure on Rh(111) surface with *W* = 18.22 Å and *d* = 2 Å. The red solid circles illustrate the bands contributed by Carbon atoms.

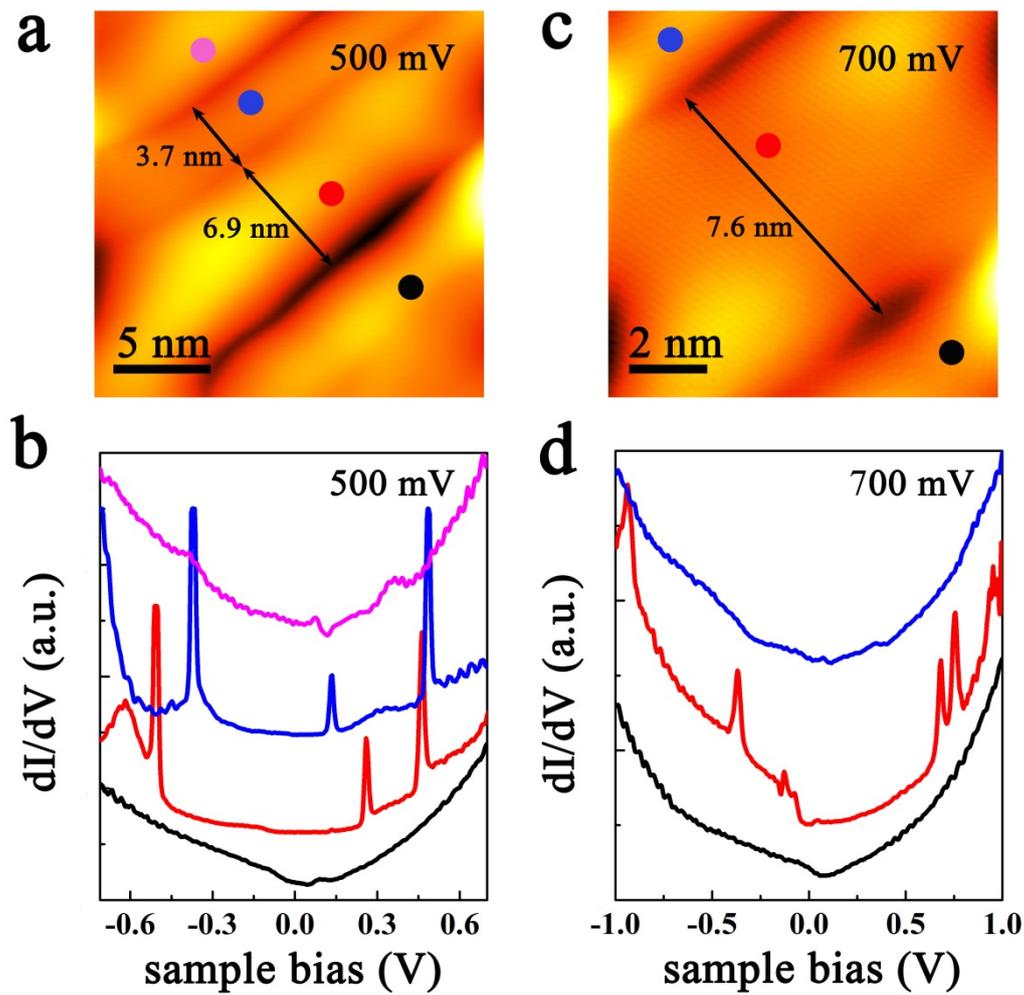

**Figure 4. Manipulation of a GNR-like structure by STM measurement. a** and **c** show two STM images of an "unstable" GNR-like structure measured subsequently. The structure of the GNR-like structure changes dramatically during the STM measurement. **b** and **d** show representative STS spectra recorded at different positions in **a** and **c**, respectively.